\documentclass[12pt,titlepage]{article}

\usepackage{graphicx}
\usepackage[utf8]{inputenc}
\usepackage[dvipsnames]{xcolor}
\usepackage{nicefrac}
\usepackage{amsthm}
\usepackage{comment}
\usepackage{subfigure}
\usepackage{mathtools}

\usepackage[
  style=ext-authoryear-comp, 
  maxcitenames=3,
  maxbibnames=99,
  uniquename=false,
  uniquelist=false,
  doi=false, 
  url=false, 
  isbn=false, 
  dashed=false,
  eprint=false
]{biblatex}

\DeclareInnerCiteDelims{textcite}{\bibopenbracket}{\bibclosebracket}

\DeclareOuterCiteDelims{parencite}{\bibopenbracket}{\bibclosebracket}

\renewbibmacro*{date+extradate}{%
  \iffieldundef{labelyear}
    {}
    {\printtext[parens]{%
       \printlabeldateextra}}}

\renewbibmacro{in:}{%
  \ifentrytype{article}
    {}
    {\printtext{\bibstring{in}\intitlepunct}}}

\DeclareDelimFormat[bib]{nametitledelim}{\addcolon\space}

\DeclareDelimFormat{nameyeardelim}{\addcomma\space}

\let\cite\textcite

\usepackage[colorlinks=true, urlcolor=blue, linkcolor=blue, citecolor=red]{hyperref}
\usepackage[noabbrev,nameinlink,capitalise]{cleveref}
\usepackage{multirow,array}
\usepackage[margin=1in]{geometry}
\usepackage{setspace}
\usepackage{url}
\usepackage{authblk}
\usepackage{paralist}
\usepackage{bm}
\usepackage{bbm}
\usepackage{amsmath,amssymb}
\usepackage{tikz}
\usepackage{dirtytalk}
\usepackage{pgfplots}
\newtheorem{theorem}{Theorem}[section]
\newtheorem{proposition}{Proposition}[section]

\newtheorem{lemma}{Lemma}[section]

\def\Re{\mathbb{R}}

\newcommand{\df}[1]{\underline{#1}}

\DeclareMathOperator{\E}{\mathbb{E}}

\allowdisplaybreaks

\theoremstyle{definition}
\newtheorem{definition}{Definition}[section]
\newtheorem{example}{Example}[section]
\title{Winner's bliss: interdependent-value auctions with horizontal uncertainty\thanks{The problem of analyzing auctions under winners' bliss was first suggested to us by Matt Shum. We are very grateful to Matt for prompting us to work on this model. We are also grateful for the reviewers and audience at EC'26, where the paper has been published as an extended abstract.}}
\author{
Federico Echenique\\
University of California, Berkeley
\and
Yu-Ting Ho\\
University of California, Berkeley
}

\begin{document}

\maketitle

\begin{abstract}
    We analyze auctions with interdependent values and horizontally differentiated bidders, where good news for one bidder is bad news for the other bidders. This generates a winner's bliss, reversing the well-known winner's curse: winning means buying an object that is more valuable than a na\"{\i}ve estimate suggests. Our main result is about the English auction. We introduce the ``own effect property,'' generalizing the single-crossing condition, to prove the existence of an efficient ex-post equilibrium.  Regarding revenue, we show that with affiliated signals, the English auction yields no more revenue than the second-price auction, while independent signals yield the standard revenue equivalence. 
\end{abstract}

\newpage
\thispagestyle{empty}
\tableofcontents

\newpage \setcounter{page}{1}

\section{Introduction}

We study auctions with interdependent values and horizontally, rather than vertically, differentiated bidders. In such auctions, good news for one bidder is necessarily bad news for the other bidders. In contrast to the famous winner's curse phenomenon of common value auctions, our model features a \emph{winner's bliss}. A winning bidder's na\"{\i}ve inference regarding the value of the object for sale is below its expected value, conditional on winning the object.

An example is bidding for drilling rights in the US outer continental shelf.  Oil and gas companies bid for off-shore tracts based on noisy information regarding the quantity and nature of the underground deposits. Bidders are not allowed to drill wells before the auction is conducted. A tract that is desirable for an oil company may not be a good purchase for a gas company, and vice versa. This means that if an oil company's prospective survey indicates it is likely to be an oil field, then this same information is bad news for a gas company that is bidding on the same tract. See \cite{HAILE2010390} for a description of oil and gas tract auctions; they also provide a review of the relevant academic literature.

More generally, we may think of bidders as firms that are invested in different technological standards. The item for sale may be a better fit for one technology and not for the others. Each bidder will then have information regarding the probable fit with their own technology. If one firm learns that the object is likely valuable to them, then at the same time, it will likely be worth less to the remaining bidders.

We study different auction formats in the setting of interdependent values with horizontally differentiated bidders. Our main results concern the English auction, an auction format that is particularly interesting when there are interdependent values \parencite{krishna2003asymmetric}. When an object is for sale in an English auction, the price rises continuously, and bidders drop out sequentially. The item goes to the last bidder standing, who pays the price at which the second-to-last bidder dropped out. As bidders exit the auction, they convey information on the value of their signals to the remaining bidders.

For English auctions, we identify a condition on bidders' valuation functions, the ``own effect property,'' that ensures the existence of an ex-post equilibrium and allows us to show that this equilibrium is efficient. The own effect property generalizes the single-crossing property assumed by \cite{maskin92} to prove that the English auction is efficient in a model with two bidders. In all, our result (Section~\ref{sec:english}) provides a normative argument for using the English auction in models that satisfy our assumptions.

In addition to the English auction, we consider the three standard sealed-bid auction formats: first-price, second-price, and all-pay auctions (Section~\ref{sec:sealed}). For the sealed-bid auction formats, we need rather strong assumptions in order to make progress. Most of our results rely on the assumption of symmetric bidders. In all three auction formats, we adapt standard techniques to construct equilibrium strategies for our winner's bliss environments.

Our main result concerns efficiency, but we also study the revenue properties of the different auction formats (Section~\ref{sec:revenue}). When signals are affiliated, the English auction generates no more revenue than the second-price auction. When signals are independent, we obtain the standard revenue-equivalence conclusion. The first-price, second-price, all-pay, and English auctions all yield the same expected revenue. 

Finally, we consider the revenue comparison between the standard winner's curse setting and our model with winner's bliss. The question is conceptually complicated, as it involves comparing revenue across different markets, not across different auction formats in the same market. Nevertheless, our preliminary findings regarding this comparison suggest that revenue is higher under the winner's curse than under bliss.

The problem of analyzing horizontal uncertainty and the winners' bliss was first proposed to us by Matt Shum and is analyzed in the (independent) work of \cite{chen2026winnersblisscommonvalueauctions}. Their paper considers a two-bidder model of the first-price auction under horizontal differentiation. The results are logically unrelated to ours but serve to illustrate and motivate interest in and study of horizontal differentiation in models of auctions. 

Most auction settings probably involve a mix of vertical and horizontal differentiation. The literature has focused on the case of pure vertical differentiation, but most real-world markets probably involve some level of agent-specificity that leads to a component of horizontal differentiation. In the paper, we focus on a model of pure horizontal differentiation, but we hope to spark interest in hybrid models that feature and combine both situations.

\section{Model}\label{sec:model}
    
A finite set of bidders \( N = \{1, 2, \dots, n\} \) participate in an auction for a single indivisible object. Before submitting a bid, each bidder receives a signal that is related to the value of the object. The bidders' preferences are interdependent: the value of the object to any one of them depends on the signals that have been received by all the bidders. 

The literature on auctions with interdependent valuations has focused on the case when an increase in any individual bidder's signal translates into an increase in any other bidders' valuation for the object. Our model deviates from this assumption and focuses instead on when the only signal that has a positive influence on a valuation is the bidders' own signal. The other bidders' signals have a negative influence. This setting models a scenario where good news for other bidders is bad news for any individual bidder. Instead of the usual  winner's curse, here we have a ``winner's bliss.''

Formally, each bidder \( i \), where \( 1 \leq i \leq n \), receives a private signal, which is a realization of a random variable \( X_i \in [0, 1] \). It is assumed that each bidder \( i \)'s value of the object, denoted by \( V_i \), can be expressed as a function of all bidders' signals. That is,
\[
V_i = v_i(X_1, X_2, \dots, X_n),
\]
where \( v_i:[0,1]^n \to \mathbb{R} \) is strictly monotonically increasing in \( X_i \), but strictly monotonically decreasing in \( X_j \) for all \( j \neq i \). Furthermore, \( v_i \) is assumed to be twice continuously differentiable (the restriction to $[0,1]^n$ of a $C^2$ function on an open superset).

To illustrate what we mean by winner's bliss, suppose (momentarily) that bidders have symmetric value functions and employ the same strategy; suppose that this strategy is an increasing function of the expected value of the good, conditioned on the signal each bidder receives. This means that \( \E[V_i \mid X_i = x_i] \). When bidder \( i \) wins the auction, we know that
\[
\E[V_i \mid X_i = x_i] < \E[V_i \mid X_i = x_i, Y_i < x_i],
\]
where \( Y_i \) denotes the highest signal received by bidders other than bidder \( i \). In words, this indicates that the winner of the auction underestimates the value of the good: a winner's bliss in opposition to the traditional winner's curse.

Under the winner’s bliss effect, bidders may, in principle, be motivated to raise their bids to improve their likelihood of winning the auction, assuming their bids are based solely on their own signals. Naturally, winner's bliss arises only when bidders misestimate the item's expected value. In equilibrium, bidders correctly condition on the event of winning the object, and the winner's bliss (like the winner's curse) is not a problem.

We have introduced the setting in this section. Next, we proceed by discussing two examples that illustrate the kind of problems we wish to capture and that show how the assumptions we make in our model are operationalized in specific instances. 

In the rest of the paper, we investigate the efficiency and revenue properties of different auction formats under our assumptions. We focus on the English auction, which we view as particularly interesting for our setting, as it discloses information during the auction process. We are not going to write much more in the way of motivation. The English auction is a very common format, and it has been studied extensively in the literature. More fundamentally, compared to other auction formats, the English auction allows us to derive more general and interesting results.

\subsection{Examples}\label{sec:examples}

\subsubsection{Linear valuations}
Let \( \theta \in \mathbb{R}^d_{+} \) be a vector representing an item or a resource up for auction. The item is characterized by \( d \) attributes (where typically we would have \( d \) being large and exceeding the number of bidders). Each bidder \( i \) has a linear valuation \( w_i \), such that the value of the item \( \theta \) for bidder \( i \) is given by \( \theta \cdot w_i \).

As a concrete example, the auction could involve a procurement contract, such as delivering services to a local government. The attributes might include the quality of the service, the geographical locations where the service needs to be provided, and so on. Each bidder has a distinct valuation for these attributes, reflecting a complex pattern of specialization.

How is $w_i$ connected to bidders' information or signals? We make some assumptions that capture the idea of horizontal uncertainty. In particular, we assume that $w_i$ is the sum of two terms: a constant baseline value common to all bidders and a term that reflects bidders'information: each bidder \( i \) is identified by a vector \( \alpha_i \in \mathbb{R}^d \), such that the ``ideal'' item for bidder \( i \) is \( \bar{\theta} + \alpha_i \), where \( \bar{\theta} \in \mathbb{R}^d \) is a constant common value component. The vector \( \alpha_i \) captures the specific preferences of each bidder \( i \) over the possible attributes of the item for sale.

Now assume that agent \( i \)'s valuation vector is their ideal item: \( w_i = \bar{\theta} + \alpha_i \). Furthermore, suppose (crucially) that \( \alpha_i \cdot w_i > 0 \) and \( \alpha_i \cdot w_j < 0 \) for any \( i \) and \( j \in N \) such that \( j \neq i \). For example, when \( d = 2 \), let \( \bar{\theta} = (1, 1) \), \( \alpha_1 = (1, -1/2) \), and \( \alpha_2 = (-1/2, 1) \). Then, bidder 1's and bidder 2's ideal items are \( (2, 1/2) \) and \( (1/2, 2) \), respectively, and the dot product is \( \alpha_j \cdot w_i = -1/2 \) for \( j \neq i \).

The actual item up for auction may not coincide with any of the bidders’ ideal items. It is given by
\[
\theta = \bar{\theta} + \sum_{j=1}^n x_j \alpha_j,
\]
where the scalars \( x_j \in [0, 1] \) are randomly drawn from some distribution \( F \) on \( [0, 1] \). The interpretation of \( x_j \) is that each bidder \( j \) receives a signal disclosing partial information about how close the item up for auction is to their ideal item.

The valuation of a bidder $i$ is now given by the function $v_i$, where 
\[
v_i(x_1,\ldots,x_n) = w_i \cdot \theta = w_i \cdot \left(\bar{\theta} + \sum_{j=1}^n x_j \alpha_j \right) = w_i \cdot \bar{\theta} + \sum_{j=1}^n x_j w_i \cdot \alpha_j 
\]
Since $w_i \cdot \alpha_j < 0$ holds for all $j \neq i$, $v_i$ is monotone decreasing in $x_j$ for all $j \ne i$. We also have $\alpha_i \cdot w_i >0$. So $v_i$ is monotone increasing in $x_i$.  Observe that $v_i(x_1,\ldots,x_n)$ is maximized when $x_i=1$ and $x_j=0$ for all $j\neq i$, which corresponds to the value of the ideal object for bidder $i$, $\bar\theta + \alpha_i$.

\subsubsection{Auctions for financial assets and a binary state space}
Consider a financial asset that is up for sale through an auction; say it is a stock. Imagine that there are two types of stocks: value and growth.  
Now, one asset is up for sale and its type is unknown. Two investors are bidding for the stock, each preferring a different kind. They share a common prior over the type of stock, but one prefers to buy a growth stock and the other prefers a value stock.

Let the state space be \(S=\{V,G\}\) with prior \(\Pr(S=V)=\rho\) and \(\Pr(S=G)=1-\rho\).
Each bidder \(i\in\{1,2\}\) observes a continuous signal \(x_i\) with the support $(0,1)$.  
If \(S=V\) then \(x_i\sim F_i^V\) with density \(f_i^V\);  
if \(S=G\) then \(x_i\sim F_i^G\) with density \(f_i^G\).  
Assume that
\[
\frac{f_1^V(x_1)}{f_1^G(x_1)} \text{ is increasing in } x_1,
\qquad
\frac{f_2^V(x_2)}{f_2^G(x_2)} \text{ is decreasing in } x_2.
\]

Denote the posterior belief by $\theta(x_1,x_2)=\Pr(S=V\mid x_1,x_2).$ By Bayes’ rule, we have that 
\[
\theta(x_1,x_2)
= \frac{\Pr(S=V)\,f_1^V(x_1)f_2^V(x_2)}
{\Pr(S=V)\,f_1^V(x_1)f_2^V(x_2)+\Pr(S=G)\,f_1^G(x_1)f_2^G(x_2)}.
\]
Dividing by $\text{Pr}(S=G \mid x_1,x_2) = 1-\theta(x_1,x_2)$ gives:
\[
\frac{\theta(x_1,x_2)}{1-\theta(x_1,x_2)}
=
\frac{\Pr(S=V)}{\Pr(S=G)}
\cdot
\frac{f_1^V(x_1)}{f_1^G(x_1)}
\cdot
\frac{f_2^V(x_2)}{f_2^G(x_2)}.
\]

Define
\[
C:=\frac{\Pr(S=V)}{\Pr(S=G)},\qquad L_1(x_1):=\frac{f_1^V(x_1)}{f_1^G(x_1)},\qquad L_2(x_2):=\frac{f_2^V(x_2)}{f_2^G(x_2)}.
\]
Then
\[
\frac{\theta(x_1,x_2)}{1-\theta(x_1,x_2)}  = C\,L_1(x_1)\,L_2(x_2), \,\,\; \text{ thus } \,\,\; \theta(x_1,x_2) = \frac{C\,L_1(x_1)L_2(x_2)} {1+C\,L_1(x_1)L_2(x_2)}.
\]
Investor 1 prefers the Value state ($V$), and Investor 2 prefers the Growth state ($G$). Their expected values, $v_1$ and $v_2$, depend on the posterior belief $\theta(x_1, x_2) = \Pr(S=V \mid x_1, x_2)$ and their own private signal $x_i$ through the intrinsic value functions $V_i(x_i)$.

The value functions are defined by:
\begin{align*}
v_1(x_1, x_2) &= \theta(x_1, x_2) V_1(x_1) \\
v_2(x_1, x_2) &= (1 - \theta(x_1, x_2)) V_2(x_2)
\end{align*}
We assume that the intrinsic value functions are increasing in the signal: $V_1'(x_1) > 0$ and $V_2'(x_2) > 0$.

We return to these examples after stating our main result.

\section{English Auction}\label{sec:english}
    
Our main results concern the English auction. The auction starts at a price of zero and rises continuously. Each bidder can decide to remain in the auction or drop out (permanently) at any given time. The auction concludes when only one bidder remains. The remaining bidder wins the item and pays the price at which the last bidder drops out. See, for example, \cite{krishna2009auction}.

As the English auction proceeds and bidders drop out, information about the bidders' private signals may be inferred. Recall that $N$ denotes the set of all bidders. Let $A \subseteq N$ denote the set of bidders that remain in the auction at a particular point in time. Let $\mathbf{p}_{N \setminus A} \in \mathbb{R}_{\ge 0}^{|N \setminus A|}$ record the price at which each bidder in $N \setminus A$ dropped out.

We define bidding strategies following the approach of \textcite{krishna2003asymmetric}: A \textbf{bidding strategy} for bidder $i$ is a collection of functions: \[\beta_i^A:[0,1] \times \mathbb{R}_{\ge 0}^{|N \setminus A|} \to \mathbb{R}_{\ge 0},\] where $i \in A \subseteq N$ and $|A| > 1$. The value $\beta_i^A (x_i,\mathbf{p}_{N \setminus A})$ represents the price at which bidder $i$ plans to drop out from the auction, given that she holds signal $x_i$, and the bidders in $N \setminus A$ have dropped out at the prices $\mathbf{p}_{N \setminus A}$.

A \textbf{strategy profile} $\bm{\beta} = ((\beta^A_i)_{i \in A})_{A \subseteq N}$ is an \textbf{ex-post equilibrium} of the English auction if, for all realizations of the signal profile $\mathbf{x} = (x_1,\dots,x_n) \in [0,1]^n$, there is no bidder who can be better off by unilaterally deviating from $\bm{\beta}$. 

An ex-post equilibrium is \textbf{efficient} if, for all realizations of the signal profile $\mathbf{x} \in [0,1]^n$, the object is allocated to the bidder with the highest value. That is, for any $\mathbf{x}\in [0,1]^n$, if $i$ wins the auction then $v_i(\mathbf{x})=\max\{v_j(\mathbf{x}) \mid j\in N \}$.

To obtain an efficient equilibrium in the English auction, we impose restrictions on valuations that are motivated by similar conditions in \cite{maskin92}, \cite{krishna2003asymmetric} and \cite{dubra2009english}. In particular, we require that for every bidder the positive effect of an increase in their own signal overwhelms the effect of a similar increase in the other bidders' signals. 

\begin{definition}
    A valuation $\mathbf{v} = (v_1,...,v_n)$ satisfies Own-Effect Domination (OED) if 
    \[
    \sum_{j=1}^n \frac{\partial v_i}{\partial x_j} > 0, \ \ \forall \ 1\le i \le n.
    \]

\end{definition}

In the two-bidder model of \cite{maskin92}, the OED results from a single-crossing property. With multiple bidders, we need to impose the OED directly. The OED property implies that, given a uniform increase in all the bidders' signals, the increase in the  valuation of bidder $i$ resulting from $i$'s own signal will dominate the aggregate value decrease from all the other bidders' signals. This property is essential for establishing an ex-post equilibrium. Our main existence result is as follows:

\begin{theorem}\label{thm:english}
    Suppose a valuation $\mathbf{v}$ satisfies OED and $ \max_{i \in N} v_i(\mathbf{0})< \min_{i \in N} v_i(\mathbf{1})$. Then there exists an ex post equilibrium of the English auction that is efficient.   
\end{theorem}

The equilibrium in Theorem~\ref{thm:english} is constructed as follows. Given a price $p$, we consider the system of nonlinear equations 
\begin{align*}
v_1(\mathbf{x}) & = p \\
v_2(\mathbf{x}) & = p \\
\vdots & \\
v_n(\mathbf{x}) & = p,    
\end{align*}
where the unknown is $\mathbf{x}\in[0,1]^n$.

For any price $p \in \left[ \max_{i \in N} v_i(\mathbf{0}), \min_{i \in N} v_i(\mathbf{1}) \right]$, we obtain a signal profile $\mathbf{x}$ that solves the above system using an argument based on the Poincar\'{e}–Miranda Theorem. In fact, using the OED, we can show that this solution is unique. Then we can apply the implicit function theorem to obtain a set of smooth functions $\sigma = (\sigma_1,\dots,\sigma_n)$, where $\sigma_i:  \left[ \max_{i \in N} v_i(\mathbf{0}), \min_{i \in N} v_i(\mathbf{1}) \right] \to [0,1]$, such that, for all $p \in  \left[ \max_{i \in N} v_i(\mathbf{0}), \min_{i \in N} v_i(\mathbf{1}) \right]$, $v_i(\sigma(p)) = p$, for all $i \in N$.

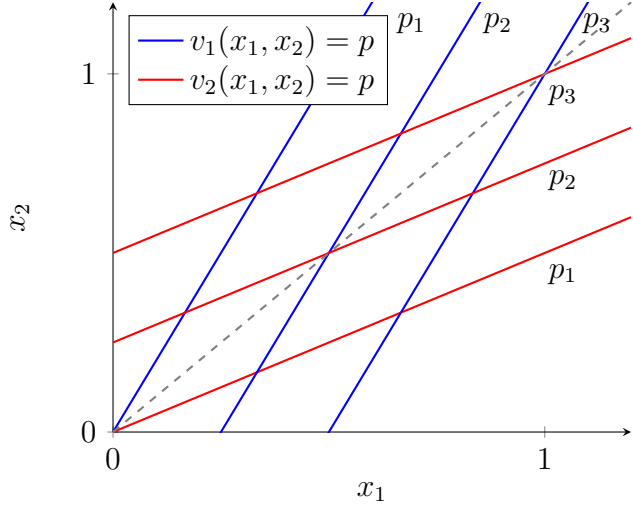
\begin{figure}
    \centering
    \begin{tikzpicture}
    \begin{axis}[
        axis lines = left,
        xlabel = {$x_1$},
        ylabel = {$x_2$},
        legend pos=north west,
        domain=0:1.2, 
        samples=100,
        ymin=0, ymax=1.2, 
        xtick={0,1}, 
        xticklabels={$0$, $1$}, 
        ytick={0,1}, 
        yticklabels={$0$, $1$},
    ]
    
    \addplot[thick, blue] {2*x};
    \addlegendentry{$v_1(x_1,x_2) = p$}
    \addplot[thick, red] {x/2};
    \addlegendentry{$v_2(x_1,x_2) = p$}

    \addplot[thick, blue] {2*x - 0.5};
    \addplot[thick, blue] {2*x - 1};

    \addplot[thick, red] {(0.5+x)/2};
    \addplot[thick, red] {(1+x)/2};
    \addplot[thick, gray, dashed] {x};

    \node at (axis cs:0.75,1.2) [anchor=north east, black] {$p_1$};

    \node at (axis cs:0.95,1.2) [anchor=north east, black] {$p_2$};

    \node at (axis cs:1.18,1.2) [anchor=north east, black] {$p_3$};

    \node at (axis cs:1.1,0.5) [anchor=north east, black] {$p_1$};

    \node at (axis cs:1.1,0.76) [anchor=north east, black] {$p_2$};

    \node at (axis cs:1.1,1) [anchor=north east, black] {$p_3$};

    \end{axis}
\end{tikzpicture}
    \caption{Let there be two bidders with value functions \( v_i(x_i, x_{-i}) = 2x_i - x_{-i} \), where the blue and red lines represent the value functions of each bidder, respectively. Given a price \( p \), the intersections of the two value functions will correspond to the function \( \sigma \) (represented by the gray-dotted line). Note that $p_1<p_2<p_3$: we can observe that \( \sigma \) is increasing in \( p \).
}
    \label{fig:sigma}
\end{figure}

An illustration of how we obtain the $\sigma$ functions is presented in \cref{fig:sigma}. Note how the level curves of the $v_i$ functions cross at larger points when $p$ increases ($p_1<p_2<p_3$). This crucial monotonicity property is ensured by the OED.\footnote{In fact the result may be seen as an instance of a more general abstract result on monotone comparative statics: see especially the working paper version of \cite{dubra2009english}, which uses the monotone comparative statics result to prove the Stolper-Samuelson theorem of international trade. }

After deriving $\sigma$, we may define a bidding strategy \(\beta_i^N(x_i) = p\) such that \(\sigma_i(p) = x_i\) for all \(1 \leq i \leq n\) at the beginning of the auction. That is, as long as $\sigma_i$ is invertible, \(\beta_i^N(x_i) = \sigma_i^{-1}(x_i)\), which implies that bidder \(i\) plans to drop out at price \(\beta_i^N(x_i)\) if she receives the signal \(x_i\). Let $j$ be the first bidder to drop out at \(\beta_j^N(x_j)\). We may back out the signal $x_j$ and solve the following system, 
\[
v_1(x_j,\mathbf{x}_{-j}) =\dots v_{j-1}(x_j,\mathbf{x}_{-j}) = v_{j+1}(x_j,\mathbf{x}_{-j})  \dots = v_n(x_j,\mathbf{x}_{-j}) = p,
\]
which will induce another $\sigma$, and the bidding strategy $\beta_i^{N \setminus \{j\}}$ can be established. This procedure ensures that the strategy profile $\bm{\beta}$ is uniquely and unambiguously determined.

To show that these functions $\sigma$ define an equilibrium, the results in \cite{krishna2003asymmetric} require that they be monotonically non-decreasing.  In our proof of Theorem~\ref{thm:english}, we use the OED to establish this property. In fact, our next example illustrates that in the absence of the OED property, the resulting strategy profile $\bm{\beta}$ fails to be an ex-post equilibrium. 

\begin{example}\label{nonincrease}
    Let there be two bidders $1$ and $2$. Assume that $x_1$ and $x_2$ are both drawn from $U[0,1]$, and the value function $v_1(x_1,x_2)$ and $v_2(x_1,x_2)$ are defined as $v_1(x_1,x_2) = x_1-2x_2+2$ and $v_2(x_1,x_2) = x_2-2x_1+2$. To derive the $\sigma$ function, we set $v_1(x_1,x_2) = v_2(x_1,x_2)$ and derive that $\beta_1(x) = \beta_2(x) =2 -x$.

    Assume that $x_1 = 0.5$ and $x_2 = 0.7$. If both players apply the bidding function $\beta$, then players 1 and 2 will drop out at $p_1 = \beta_1(0.5) = 1.5$ and $p_2 = \beta_2(0.7) = 1.3$, respectively. In this case, 2 will drop out first, and 1 gets the information that $x_2 = 0.7$, which implies that $v_1(0.5,0.7) = 1.1 < 1.3$. Bidder 1 would be better off by deviating from $\beta_1$ and dropping out earlier. Hence, $\bm{\beta}$ fails to constitute an equilibrium when the OED condition is violated.
\end{example}

For the last part of the theorem, the OED property also guarantees the efficiency of the equilibrium. The proof is presented in the appendix. At the end of this section, we present examples that meet our assumptions and serve to illustrate how the equilibrium is constructed.

\subsection{A linear valuation model.}
    Consider the linear model in Section~\ref{sec:examples}, where
\[
w_i=\bar\theta+\alpha_i,\qquad
\theta(\mathbf{x})=\bar\theta+\sum_{j\in N}x_j\alpha_j,
\qquad
v_i(\mathbf{x})=w_i\cdot\theta(\mathbf{x}).
\]
Maintain the restrictions $w_i\cdot\alpha_i>0$ and $w_i\cdot\alpha_j<0$ for $j\ne i$. Writing $S=\sum_{j\in N}\alpha_j$, impose the additional condition
\begin{equation}\label{eq:procurement-endpoints}
    \max_{i\in N} w_i\cdot\bar\theta
    <
    \min_{i\in N} w_i\cdot(\bar\theta+S).
\end{equation}
These restrictions imply the hypotheses of Theorem~\ref{thm:english}. Indeed, the valuation functions are affine, hence twice continuously differentiable, and
\[
    \frac{\partial v_i}{\partial x_j}=w_i\cdot\alpha_j
\]
has the required sign. Moreover, \eqref{eq:procurement-endpoints} implies, for every bidder $i$,
\[
    \sum_{j\in N}\frac{\partial v_i}{\partial x_j} = w_i\cdot S > \max_{k\in N}w_k\cdot\bar\theta-w_i\cdot\bar\theta \ge 0,
\]
so OED holds. Finally,
\[
    v_i(\mathbf{0})=w_i\cdot\bar\theta, \qquad v_i(\mathbf{1})=w_i\cdot(\bar\theta+S),
\]
and \eqref{eq:procurement-endpoints} is the theorem's endpoint condition. Thus the English auction admits an efficient ex-post equilibrium.

The additional restriction has a particularly simple interpretation when baseline valuations $w_i\cdot\bar\theta$ are equal across bidders: it requires $w_i\cdot S>0$ for every $i$. Although each individual signal harms the other bidders, a simultaneous increase in all signals must benefit every bidder.

The numerical example outlines earlier satisfies these conditions, take
\[ \bar\theta=(1,1),\qquad \alpha_1=(1,-1/2),\qquad \alpha_2=(-1/2,1).
\]
Then $w_1=(2,1/2)$, $w_2=(1/2,2)$, and
\[
    v_1(\mathbf{x})=\frac52+\frac94x_1-\frac12x_2,
    \qquad
    v_2(\mathbf{x})=\frac52+\frac94x_2-\frac12x_1.
\]
The own-signal derivatives are $9/4$, the cross-signal derivatives are $-1/2$, and their row sums are $7/4>0$. Furthermore,
\[
    \max_i v_i(\mathbf{0})=\frac52  <\frac{17}{4}=\min_i v_i(\mathbf{1}).
\]
Both coordinates of $\theta(\mathbf{x})$ are at least $1/2$ throughout $[0,1]^2$, so the item also satisfies the maintained nonnegativity restriction on its attributes.

\subsection{A financial asset with a binary state space.}

We return to the value/growth stock example that we described in Section~\ref{sec:examples}. It will be particularly relevant for the comparison exercise in Section~\ref{sec:horizontal-vertical-comparison}.

We first determine when the OED will be satisfied. Define the sum of the partial derivatives as $\Sigma v_i = \frac{\partial v_i}{\partial x_1} + \frac{\partial v_i}{\partial x_2}$, and $\Sigma \theta = \frac{\partial \theta}{\partial x_1} + \frac{\partial \theta}{\partial x_2}$ as the overall belief shift. We have
\begin{align*}
\Sigma v_1 &= \theta(x_1, x_2) V_1'(x_1) + V_1(x_1) \cdot \Sigma \theta\\
\Sigma v_2 &= (1-\theta(x_1, x_2)) V_2'(x_2) - V_2(x_2) \cdot \Sigma \theta
\end{align*}
The conditions for $\Sigma v_i > 0$ are:
\begin{align*} 
\theta(x_1, x_2) V_1'(x_1) &> - V_1(x_1) \cdot \Sigma \theta\\
(1-\theta(x_1, x_2)) V_2'(x_2) &> V_2(x_2) \cdot \Sigma \theta
\end{align*}

Note that the left hand side is the direct effect of the increase in the intrinsic value of the good, which comes from the increment of each bidder's own signal. The right hand side is the indirect effect, which comes from the shift of the posterior belief. If $\Sigma \theta$ is positive, this favors investor 1, and the inequality always holds. On the other hand, if $\Sigma \theta$ is negative, then the direct effect must compensate for the indirect effect.  Next, we assume specific functional forms and solve the equilibrium profile.

\begin{example}
    Let $\pi = 0.5$, \(C=1\) and
\[
F_1^V(x)=x^2,\qquad F_1^G(x)=1-(1-x)^2=2x-x^2,
\]
\[
F_2^V(x)=1-(1-x)^2=2x-x^2,\qquad F_2^G(x)=x^2.
\]
Their densities are
\[
f_1^V(x)=2x,\quad f_1^G(x)=2(1-x),\qquad
f_2^V(x)=2(1-x),\quad f_2^G(x)=2x.
\]
Thus,
\[
L_1(x_1)=\frac{x_1}{1-x_1},
\qquad
L_2(x_2)=\frac{1-x_2}{x_2}.
\]
With \(C=1\), for each pair of $(x_1,x_2) \in (0,1)^2$, the posterior becomes
\[
\theta(x_1,x_2)
=
\frac{\frac{x_1}{1-x_1}\cdot\frac{1-x_2}{x_2}}
{1+\frac{x_1}{1-x_1}\cdot\frac{1-x_2}{x_2}}
=
\frac{x_1(1-x_2)}{x_1(1-x_2)+x_2(1-x_1)}.
\]
\end{example}

Consider the following exponential function as a concrete example. Let $V_1(x_1) = exp \{ \alpha x_1 \}$ and $V_2(x_2) = exp \{ \alpha x_2 \}$. For the condition that $\Sigma v_1 > 0$, we must have
\begin{align*}
    & \theta(x_1, x_2) V_1'(x_1) > - V_1(x_1) \cdot \Sigma \theta \implies  \frac{V_1(x_1)}{V'_1(x_1)} > -\frac{\Sigma \theta}{\theta(x_1,x_2)}\\
    \implies & \alpha > \max_{x_1, x_2} \left| \frac{1}{\theta(x_1, x_2)} \left( \frac{\partial \theta}{\partial x_1} + \frac{\partial \theta}{\partial x_2} \right) \right|
\end{align*}

Similarly, we can also derive the condition that $\Sigma v_2 > 0$ as follows:

\begin{align*}
    \alpha > \max_{x_1, x_2} \left| \frac{1}{1-\theta(x_1, x_2)} \left( \frac{\partial \theta}{\partial x_1} + \frac{\partial \theta}{\partial x_2} \right) \right|
\end{align*}

Since $x_i \in (0,1)$, by selecting sufficiently large positive constants $\alpha$, the sum of the partial derivatives is strictly positive for both investors across all possible signal combinations, which guarantees the OED condition. Now, we are ready to solve for the equilibrium. We first solve the system
\begin{align*}
    v_1(x_1,x_2) = v_2(x_1,x_2) = p
\end{align*}

We have 
\begin{align*}
    v_1(x_1,x_2) & = \theta(x_1,x_2) exp\{\alpha x_1\} =  p \implies  \theta(x_1,x_2) = p \cdot exp\{-\alpha x_1\}\\
    v_2(x_1,x_2) & = [1-\theta(x_1,x_2)] exp\{\alpha x_2\} =  p  \implies  [1- \theta(x_1,x_2)] = p \cdot exp\{- \alpha x_2\}\\
    \implies & \theta(x_1,x_2) + [1-\theta(x_1,x_2)] = p \cdot (exp\{-\alpha x_1\} + exp\{- \alpha x_2\})\\
    \implies & p(x_1,x_2) = \frac{1}{exp\{-\alpha x_1\} + exp\{- \alpha x_2\}}
\end{align*}
Thus, 
\begin{align*}
    \frac{\partial{p}}{\partial{x_i}} = \frac{\alpha exp \{-\alpha x_i \}}{(exp\{-\alpha x_1\} + exp\{- \alpha x_2\})^2} > 0,
\end{align*}
which proves that $p$ is increasing in $x_1$ and $x_2$. We can obtain the interval of $p(x_1,x_2)$ while $(x_1,x_2) \in (0,1)^2$. We have

\begin{align*}
    &p(0,0) = \frac{1}{exp\{-\alpha \cdot 0\} + exp\{- \alpha \cdot 0\}} = \frac{1}{2}\\
    &p(1,1) = \frac{1}{exp\{-\alpha\cdot 1\} + exp\{- \alpha \cdot 1\}} = \frac{1}{2exp\{-\alpha\}} \\
    \implies &p(x_1,x_2) \in \left(\frac{1}{2},\frac{exp\{\alpha\}}{2 } \right)
\end{align*}

Now, we start to derive the bidding strategy $\beta_i = \sigma^{-1}_i$. First, consider the system
\begin{align*}
    & v_1(x_1,x_2) = v_2(x_1,x_2) \implies \frac{\theta(x_1,x_2)}{1-\theta(x_1,x_2)} =\frac{V_2(x_2)}{V_1(x_1)}
    \implies \frac{x_1(1-x_2)}{x_2(1-x_1)} = exp \{ \alpha (x_2-x_1)\}
\end{align*}

Note that if $x_1 \ne x_2$, then the equality will not hold. To see that, we assume that $x_1 > x_2$ without loss of generality; then we have the left-hand side larger than $1$. However, $x_2 -x_1$ is negative, which implies that the right-hand side is less than 1. Hence, we cannot have $x_1 > x_2$ while the equality holds. Given that $x_1 = x_2$,
\begin{align*}
    p = \frac{1}{2 exp \{ -\alpha x_i \}} \ \ \forall \ i =1,2
    \implies & \sigma(p) = \left( \frac{1}{\alpha} \cdot \ln(2p),  \frac{1}{\alpha} \cdot \ln(2p) \right)\\
    \implies &\beta_i(x_i) = \frac{1}{2} exp\{\alpha x_i\},
\end{align*}
which constitutes an efficient ex-post equilibrium under the English auction. The stock will be sold within the price range $ \left(\frac{1}{2},\frac{exp\{\alpha\}}{2 } \right)$.

\section{Sealed-bid Auctions}\label{sec:sealed}
    
We turn to auction formats beyond the English auction. Relying on standard techniques (see, for example, \cite{krishna2009auction} and \cite{krishna1997analysis}), but adapted out the case of horizontal differentiation and winners' curse, we obtain characterizations of equilibrium and some basic revenue comparisons. The model is less tractable than under the English auction, and our results are weaker. In particular, we assume, for the most part, that bidders are symmetric. First, we consider the second-price auction (Section~\ref{sec:secondprice}), which, analogous to the English auction, requires no assumptions about the distribution of signals. The reason is that a bidder's payment in these signals is independent of their own signal. Then we turn to the first-price auction (Section~\ref{sec:firstprice}) in which we shall have to restrict the distribution of signals.

We start by assuming that valuations are symmetric across buyers.

\begin{definition}
A profile of valuation functions $\mathbf{v}$ is \df{symmetric} if there exists a function $v:[0,1]^N\to\Re_+$ that is symmetric in its last $n-1$ components, and for which
    \[
v_i(\mathbf{x}) = v(x_i,\mathbf{x}_{-i}) \ \forall \ 1 \le i \le n.
\]
\end{definition}

The assumption that $v$ is symmetric in its last $n-1$ components means that it is invariant to permutations of the last $n-1$ components. For example, suppose that there are three bidders with symmetric valuations and where the relevant valuation is \( v \). Then we have that \( v(a, b, c) = v(a, c, b) \). 

The following analysis is based on the assumption that value functions are symmetric among all bidders.

Before presenting our results, we reintroduce some key notation. Let \( X_i \) denote the random variable representing the signal received by bidder \( i \), and denote the profile $(X_1,\dots,X_n)$ by $\mathbf{X}$. Let \( Y_k \) denote the random variable representing the \( k \)-th highest signal received among all bidders $j\neq i$. In particular, when $i$ is understood from context, $Y_1=\max\{X_j \mid j\neq i \}$. 

We define the function \( u \) by
\[
u(x, y) = \mathbb{E}[v(\mathbf{X})| X_i = x, Y_1 = y],
\]
which represents the expected value of the object to bidder \( i \), given that they receive the signal \( x_i = x \) and the highest signal among all other bidders, \( y_1 \), is \( y \).  Let $g(\cdot|x)$ be the probability distribution of $Y_1$ conditional on $X_i = x$.

With these definitions in place, we are now ready to present our results.

\subsection{Second-price Auction}\label{sec:secondprice}

\begin{proposition}\label{prop:secondprice}
    The symmetric bidding strategy $\beta^{\mathrm{II}}(x) = u(x,x)$ constitutes an equilibrium under the sealed-bid second-price auction.\end{proposition}
    \begin{proof}
        Suppose that all other bidders $j \ne i$ follow the strategy $\beta_j(x) =  u(x,x)$. Bidder $i$'s expected payoff given that she receives the signal $x$ and bids the amount $p_i$ is
\begin{align*}
    \Pi(p_i,x) & = \int_0^{\beta^{-1}(p_i)}(u(x,y) - \beta(y))g(y|x)dy \\
    & = \int_0^{\beta^{-1}(p_i)}(u(x,y) - u(y,y))g(y|x)dy   
\end{align*}
This formulation follows from the fact that bidder $i$ will only win the object if $\beta(y) < p_i$, which implies that $y < \beta^{-1}(p_i)$. Hence, we take the integral from 0 to $ \beta^{-1}(p_i)$. Since we assume that $u(x,y)$ is increasing in $x$, 
bidder $i$ should outbid all the other bidders whenever $x > y$. To find the best response of bidder $i$, it is clear that she should choose $\beta^{-1}(p_i) = x$, which implies $p_i = \beta(x) = u(x,x)$.
    \end{proof}

\subsection{First-price Auction}\label{sec:firstprice}

\begin{proposition}\label{prop:firstprice}
    Given that $v(0,\mathbf{x}_{-i}) = 0$, for all $\mathbf{x}_{-i} \in \mathbb{R}^{n-1}$, and each signal $x_i$ is independently drawn from a distribution for all $1 \le i \le n$, the following symmetric bidding strategy $\beta^{\mathrm{I}}(x)$ constitutes an equilibrium under the sealed-bid first-price auction.
\[
\beta^{\mathrm{I}}(x) = \int_0^x u(y,y) d L(y)
\]
where 
\[
L(y|x) = exp\left(-\int_y^x\frac{g(t|t)}{G(t|t)}dt\right) 
\]
\end{proposition}

In the first-price auction, more constraints are in place. The additional assumption serves as a requirement to ensure that the bidding function, constructed by solving the ordinary differential equation, is increasing. A similar constraint and result apply to the all-pay auction, which is presented in the next subsection.

\subsection{All-pay Auction}
\begin{proposition}\label{prop:allpay}
    Given that $v(0,\mathbf{x}_{-i}) = 0$, for all $\mathbf{x}_{-i} \in \mathbb{R}^{n-1}$, and each signal $x_i$ is independently drawn from a distribution for all $1 \le i \le n$, the following symmetric bidding strategy $\beta^A(x)$ constitutes an equilibrium under the sealed-bid all-pay auction.
\[
\beta^A(x) =  \int_0^x u(y,y) g(y|x) d y
\]
\end{proposition}

As we move from the English auction to different formats of sealed-bid auctions, less information is disclosed to individual bidders, and more constraints are required to establish equilibrium. One restriction in the sealed-bid auction is that the value functions must be symmetric among all bidders. In the following subsection, we discuss the difficulties that arise in establishing equilibrium without the symmetry assumption. Some restricted results are also provided.

\subsection{Asymmetric Valuations in Sealed-bid Second-price Auction}
In this subsection, we analyze the sealed-bid second-price auction without the symmetry assumption. By \cref{lemma:implicit}, we know that, under certain assumptions, there exists a set of continuous functions $\bm{\sigma}_N = (\sigma_i)_{i\in N}$, where $\sigma_i:  \left[ \max_{i \in N} v_i(\mathbf{0}), \min_{i \in N} v_i(\mathbf{1}) \right] \to [0,1]$, such that $v_i(\bm{\sigma}_N(p)) = p$,  for all $p \in  \left[ \max_{i \in N} v_i(\mathbf{0}), \min_{i \in N} v_i(\mathbf{1}) \right]$. Hence, we can define a strategy profile $\bm{\beta}^{II} = (\beta_i)_{i \in N}$, where the bidding function of each bidder $i$, $\beta^{II}_i(x_i) = \sigma ^{-1}_i(x_i)$, is well-defined. We are going to show that $\bm\beta^{II}$ constitutes an efficient equilibrium under the sealed-bid second-price auction, when there are only two bidders. 

\begin{proposition}\label{prop:twobidder}
Suppose that there are two bidders, that the valuation $\mathbf{v}$ satisfies OED and $ \max_{i \in \{1,2\}} v_i(\mathbf{0})< \min_{i \in \{1,2\}} v_i(\mathbf{1})$. Then $\bm{\beta}^{II}$ constitutes an efficient equilibrium under the sealed-bid second-price auction.
\end{proposition}

Proposition~\ref{prop:twobidder} does not extend to the $n$ bidders case. Consider the following example,

\begin{example}
    There are three bidders, such that $v_1(x_1,x_2,x_3) = 3x_1 -x_2 -x_3$,  $v_2(x_1,x_2,x_3) = x_2 $, and 
    $v_3(x_1,x_2,x_3) = x_3 $. Each bidder $i$ gets the signal $x_i \sim U[0,1]$. Assume that bidders 2 and 3 follow the bidding strategy $\beta_j = \sigma_j^{-1}$. Then we know that $\beta_2(x_2)  = x_2$ and $\beta_3(x_3) =  x_3$. Bidder 1's expected payoff given that she receives the signal $x_1$ and bids the amount $p_1$ is
    \begin{align*}
        & \int_0^{p_1}\int_0^{x_2} v_1(x_1,x_2,x_3) - \beta_2(x_2) \ dx_3 d x_2 + \int_0^{p_1}\int_0^{x_3} v_1(x_1,x_2,x_3) - \beta_3(x_3) \ dx_2 d x_3 \\
        = & 2\int_0^{p_1} \int_0^{x_2} 3x_1 - 2x_2 - x_3 \ dx_3 dx_2 \\
        = & 3x_1p_1^2 - \frac{5}{3}p_1^3
    \end{align*}
    By taking the first order condition, we derive $p_1 = \frac{6}{5}x_1$, which is different from $\beta_1(x_1) = x_1$.
\end{example}

\section{Revenue Analysis}\label{sec:revenue}
    
Our main result focuses on efficiency as a basic normative property. This was the concern in the original work of \cite{maskin92}, which focused on auctions for a public resource. An auctioneer may, of course, also be concerned with revenue. In this section, we examine revenue from two perspectives. First, we compare revenue across auction formats. Second, we compare revenue under horizontal and vertical uncertainty.

\subsection{Revenue comparison across auction formats}

Our equilibrium results for the first-price and all-pay auctions assume independently drawn signals. Under this assumption, the standard revenue equivalence result applies: the English auction and the three sealed-bid auctions discussed in \cref{sec:sealed} generate the same expected revenue, since each allocates the object to the bidder with the highest signal.

To go beyond the standard revenue equivalence result, we compare the English and second-price auctions when signals are affiliated. The following proposition shows how their expected revenues differ.

\begin{proposition} Suppose that bidders' signals are affiliated. Then the English auction will generate weakly less expected revenue than the sealed-bid second-price auction.
\end{proposition}
    \begin{proof}
        Without loss of generality, let $1$ be the winner of the auction. Assume that all signals are affiliated; then, for all $x>y$, we have
\begin{align*}
    u(y,y) & = \E[v(x_1,y_1,\dots,y_{n-1})|x_1 = y, y_1 = y] \\ 
    & = \E[v(y_1,y_1,\dots,y_{n-1})|x_1 = y, y_1 = y] \\
    & \ge \E[v(y_1,y_1,\dots,y_{n-1})|x_1 = x, y_1 = y]
\end{align*}

The last inequality follows from the fact that all the signals are affiliated and the value function is decreasing in the last $n-1$ components. The expected revenue in a second-price auction can be written as:
\begin{align}
    \E[R^{\mathrm{II}}] & = \E[\beta^{\mathrm{II}}(y_1)|x_1 > y_1 ] \nonumber \\
    & = \E[u(y_1,y_1)|x_1 > y_1 ] \nonumber \\
    & \ge \E[\E[v(y_1,y_1,\dots,y_{n-1})|x_1 = x, y_1 = y]|x_1 > y_1 ] \label{rev} \\
    & = \E[v(y_1,y_1,\dots,y_{n-1})|x_1 > y_1 ] \nonumber 
\end{align}
Note that $\E[v(y_1,y_1,\dots,y_{n-1})|x_1 > y_1]$ is exactly the expected revenue of the English auction given that $1$ is the winner of the auction.
    \end{proof}

Note that, for the purpose of establishing equilibrium under the second-price auction, we have assumed that the value functions are symmetric, which is not required for establishing equilibrium under the English auction. In contrast to the result of standard setting, we have shown that the English auction generates weakly less revenue, rather than more, under our framework. It is also worth mentioning that independence is a special case of affiliation. If signals are independent, the equality in \eqref{rev} will hold, which coincides with the standard revenue equivalence result.

\subsection{Horizontal and vertical uncertainty: revenue and welfare}
\label{sec:horizontal-vertical-comparison}

How does horizontal uncertainty affect the seller's revenue relative to the welfare generated by the auction? Comparing different valuation environments requires a normalization. We consider two models that give each bidder the same joint distribution of her own signal and her state-contingent value. The models differ in whether the bidders prefer the same state or different states. In this comparison, horizontal uncertainty generates greater welfare, but the seller receives a strictly smaller fraction of that welfare. The ranking of revenue itself can go either way.

There are two bidders and two states, $S\in\{V,G\}$, with $\Pr(S=V)=\Pr(S=G)=1/2$. Each bidder has a preferred state. Conditional on the state, signals are independent and take values in $(0,1)$. A bidder's signal has density
\[
 \begin{aligned}
 f^+(x)&=2x &&\text{in her preferred state},\\
 f^-(x)&=2(1-x) &&\text{in the other state}.
 \end{aligned}
\]
Her state-contingent value is $e^{\alpha x}$ in her preferred state and zero otherwise, where $\alpha>0$. Thus a higher signal is favorable news about her preferred state and also increases her value conditional on that state.

In the \emph{horizontal} model, denoted by $H$, bidder 1 prefers $V$ and bidder 2 prefers $G$. In the \emph{vertical} model, denoted by $C$, both bidders prefer $V$. In both models, each bidder's signal is marginally uniform, and the joint distribution of her own signal and her state-contingent value is the same. The joint distribution across bidders changes with the alignment of their preferences. The balanced prior is part of this normalization.

Signals in each model are oriented toward the bidder's preferred state. In particular, when bidder 2's preferred state changes from $G$ to $V$, her favorable signal changes as well. In the original horizontal signal coordinates, her intrinsic value in the vertical model would therefore be $e^{\alpha(1-x_2)}$. Holding it equal to $e^{\alpha x_2}$ instead would define a different benchmark, in which bidder 2's signal need not increase her expected value.

The posterior probabilities of state $V$ are
\[
 \begin{aligned}
 \theta_H(x_1,x_2)
 &=\frac{x_1(1-x_2)}{x_1(1-x_2)+(1-x_1)x_2},\\
 \theta_C(x_1,x_2)
 &=\frac{x_1x_2}{x_1x_2+(1-x_1)(1-x_2)}.
 \end{aligned}
\]
The valuations conditional on both signals are consequently
\[
 v_1^H=\theta_H e^{\alpha x_1},\qquad
 v_2^H=(1-\theta_H)e^{\alpha x_2},\qquad
 v_i^C=\theta_C e^{\alpha x_i}\quad(i=1,2).
\]
The horizontal model has negative cross-signal effects, while the vertical model has positive cross-signal effects. Both models have strictly positive own-signal effects.

Let $R_K(\alpha)$ denote expected seller revenue in the equilibrium specified below, and let
\[
 \begin{aligned}
 W_K(\alpha)
 &=\mathbb{E}_K\!\left[\max\{v_1^K(X_1,X_2),v_2^K(X_1,X_2)\}\right],\\
 \eta_K(\alpha)&=\frac{R_K(\alpha)}{W_K(\alpha)},
 \qquad K\in\{H,C\}.
 \end{aligned}
\]
Welfare is the expected value of the object to its recipient; transfers between the bidders and the seller do not affect it. Because the equilibria below are efficient, $W_K$ is also their expected total surplus. The extraction ratio $\eta_K$ is a ratio of expectations.

\begin{proposition}\label{prop:horizontal-vertical-extraction}
For every $\alpha>0$, the English auction has efficient ex-post
equilibria with dropout strategies
\[
 \beta_H(x)=\frac12 e^{\alpha x},\qquad
 \beta_C(x)=\frac{x^2}{x^2+(1-x)^2}e^{\alpha x}.
\]
The associated welfare and revenue-extraction ratios satisfy
\[
 W_H(\alpha)>W_C(\alpha),\qquad
 \eta_H(\alpha)<\eta_C(\alpha).
\]
Nevertheless, there is a unique $\alpha^*>0$ such that
\[
 \begin{cases}
 R_H(\alpha)>R_C(\alpha),&0<\alpha<\alpha^*,\\
 R_H(\alpha)=R_C(\alpha),&\alpha=\alpha^*,\\
 R_H(\alpha)<R_C(\alpha),&\alpha>\alpha^*.
 \end{cases}
\]
Numerically, $\alpha^*\simeq1.2891$.
\end{proposition}

\begin{proof}
We first verify equilibrium and efficiency, then derive the welfare and revenue formulas and compare them.

\emph{Equilibrium and efficiency.}
In each model, valuations are equal precisely when $x_1=x_2$.
For the horizontal model, this follows from
\[
 \frac{v_1^H(x_1,x_2)}{v_2^H(x_1,x_2)}
 =\frac{x_1(1-x_2)}{x_2(1-x_1)}e^{\alpha(x_1-x_2)};
\]
for the vertical model, their ratio is $e^{\alpha(x_1-x_2)}$.
Thus the bidder with the higher signal has the higher valuation in
either model. Evaluating valuations on the diagonal gives the stated
dropout strategies, both of which are strictly increasing.

Fix a model $K$, a bidder $i$, and the opponent's signal $y$. If bidder
$i$ wins, she pays $\beta_K(y)=v_i^K(y,y)$. Her valuation is strictly
increasing in her own signal, so
\[
 v_i^K(x,y)-\beta_K(y)
 \begin{cases}
 >0,&x>y,\\
 =0,&x=y,\\
 <0,&x<y.
 \end{cases}
\]
The proposed strategy wins exactly when doing so is profitable. This establishes ex-post equilibrium. At equal signals either bidder may receive the object, with zero net surplus for the winner. The allocation is efficient in all cases. This verification applies to every $\alpha>0$ and does not require an appeal to OED.

\emph{Welfare and revenue formulas.}
The unconditional joint signal densities are
\[
 \begin{aligned}
 g_H(x_1,x_2)&=2\{x_1(1-x_2)+(1-x_1)x_2\},\\
 g_C(x_1,x_2)&=2\{x_1x_2+(1-x_1)(1-x_2)\}.
 \end{aligned}
\]
Multiplication by the appropriate posterior eliminates its denominator.
In the horizontal model, symmetry and efficiency therefore give
\[
 W_H(\alpha)
 =4\int_0^1\int_0^{x_1}
 x_1(1-x_2)e^{\alpha x_1}\,dx_2\,dx_1
 =\int_0^1 e^{\alpha t}(4t^2-2t^3)\,dt.
\]
In the vertical model,
\[
 W_C(\alpha)
 =\frac12\mathbb{E}\!\left[e^{\alpha\max\{X_1,X_2\}}\mid S=V\right]
 =2\int_0^1 t^3e^{\alpha t}\,dt,
\]
because the conditional distribution of the maximum is $t^4$.
Consequently,
\begin{equation}\label{eq:comparison-welfare-gap}
 W_H(\alpha)-W_C(\alpha)
 =4\int_0^1t^2(1-t)e^{\alpha t}\,dt>0.
\end{equation}

Let $M=\min\{X_1,X_2\}$. In the horizontal model,
\[
 \Pr_H(M>t)=(1-t^2)(1-t)^2,
 \qquad f_{M,H}(t)=2(1-t)^2(1+2t).
\]
In the vertical model,
\[
 \Pr_C(M>t)=\frac12(1-t^2)^2+\frac12(1-t)^4,
 \qquad f_{M,C}(t)=2(1-t)(1-t+2t^2).
\]
The price is the losing bid, $\beta_K(M)$. Write
\[
 \begin{aligned}
 d(t)&=t^2+(1-t)^2,\\
 r_H(t)&=(1-t)^2(1+2t),\\
 r_C(t)&=\frac{2t^2(1-t)(1-t+2t^2)}{d(t)}.
 \end{aligned}
\]
Then
\begin{equation}\label{eq:comparison-revenue-integrals}
 R_H(\alpha)=\int_0^1e^{\alpha t}r_H(t)\,dt,
 \qquad
 R_C(\alpha)=\int_0^1e^{\alpha t}r_C(t)\,dt.
\end{equation}

\emph{Revenue relative to welfare.}
Define
\[
 J(\alpha)=4\int_0^1t^2(1-t)e^{\alpha t}\,dt
 =W_H(\alpha)-W_C(\alpha).
\]
We first establish $R_C(\alpha)>J(\alpha)$ for $\alpha>0$.
Indeed,
\[
 r_C(t)-4t^2(1-t)
 =\frac{2t^2(1-t)^2(2t-1)}{d(t)}=:q(t).
\]
The function $q$ satisfies $q(1-t)=-q(t)$ and is positive for $t>1/2$.
Pairing $t$ with $1-t$ yields
\[
 R_C(\alpha)-J(\alpha)
 =\int_{1/2}^1q(t)
 \left[e^{\alpha t}-e^{\alpha(1-t)}\right]dt>0.
\]

It remains to prove
\begin{equation}\label{eq:comparison-auxiliary-ratios}
 A(\alpha):=\frac{J(\alpha)}{R_H(\alpha)}
 > B(\alpha):=\frac{W_C(\alpha)}{W_H(\alpha)}.
\end{equation}
Both $A$ and $B$ are strictly increasing in $\alpha$. To see this,
recall that if $f/g$ is increasing, then
\[
 \frac{\int_0^1 e^{\alpha t}f(t)\,dt}
      {\int_0^1 e^{\alpha t}g(t)\,dt}
\]
is increasing: its derivative is the covariance of $t$ and $f(t)/g(t)$
under the probability density proportional to $e^{\alpha t}g(t)$.
For $A$ and $B$, the relevant ratios of integrands are, respectively,
\[
 \frac{4t^2}{(1-t)(1+2t)},\qquad
 \frac{t}{2-t},
\]
both strictly increasing on $(0,1)$. Moreover, $A(0)=2/3$ and $B(\alpha)<1$.

We use the convenient cutoff $a=3/2$ to compare these two increasing
ratios. Let $I_k(a)=\int_0^1t^ke^{at}\,dt$. Direct integration gives
\[
 4I_2(a)-5I_3(a)
 =\frac{(75/8)e^{3/2}-42}{(3/2)^4}>0
 \qquad(a=3/2),
\]
and
\[
 J(a)-R_H(a)
 =7I_2(a)-6I_3(a)-I_0(a)
 =\frac{12e^{3/2}-429/8}{(3/2)^4}>0
 \qquad(a=3/2).
\]
Both signs follow, for example, from
\[
 e^{3/2}>\sum_{k=0}^{7}\frac{(3/2)^k}{k!}
 =\frac{321193}{71680}>\frac{112}{25}>\frac{143}{32}.
\]
The first inequality in integrals is equivalent to $B(3/2)<2/3$;
the second gives $A(3/2)>1$. Hence, for $0<\alpha\le3/2$,
\[
 B(\alpha)\le B(3/2)<\frac23=A(0)<A(\alpha),
\]
while for $\alpha\ge3/2$,
\[
 B(\alpha)<1<A(3/2)\le A(\alpha).
\]
This proves \eqref{eq:comparison-auxiliary-ratios}. Combining it with
$R_C>J$ gives
\[
 \frac{R_C(\alpha)}{R_H(\alpha)}
 >\frac{J(\alpha)}{R_H(\alpha)}
 >\frac{W_C(\alpha)}{W_H(\alpha)},
\]
which is equivalent to $\eta_C(\alpha)>\eta_H(\alpha)$.

\emph{The reversal in revenue.}
Subtracting the revenue integrands gives
\[
 k(t):=r_C(t)-r_H(t)
 =\frac{(1-t)(2t-1)(4t^3-2t^2+t+1)}{d(t)}.
\]
The cubic factor is positive on $[0,1]$: it equals one at zero and its
derivative, $12t^2-4t+1$, is everywhere positive. Thus $k(t)$ has the
sign of $t-1/2$. It follows that
\[
 L(\alpha):=e^{-\alpha/2}[R_C(\alpha)-R_H(\alpha)]
 =\int_0^1 e^{\alpha(t-1/2)}k(t)\,dt
\]
is strictly increasing, since
\[
 L'(\alpha)=\int_0^1
 (t-1/2)e^{\alpha(t-1/2)}k(t)\,dt>0.
\]
The integral formulas extend continuously to $\alpha=0$. At that point,
$R_H(0)=1/2$ and $R_C(0)=J(0)=1/3$, so $L(0)=-1/6$.
At $\alpha=3/2$, the preceding argument gives
$R_C(3/2)>J(3/2)>R_H(3/2)$, so $L(3/2)>0$.
Continuity and strict monotonicity establish a unique zero
$\alpha^*\in(0,3/2)$ and the stated revenue ranking.
\end{proof}

The proposition separates the effect on the size of the surplus from
the effect on its division. Horizontal differentiation raises welfare:
in either state, one bidder values the object, whereas in the vertical
model neither bidder values it in state $G$. Nevertheless, the seller
captures a smaller fraction of welfare under horizontal uncertainty
for every $\alpha>0$. If welfare in each environment is normalized to
one by a common positive scaling of that environment's valuations,
the corresponding equilibrium revenue is therefore strictly larger in
the vertical model. In unnormalized units, revenue goes in the opposite
direction when intrinsic values vary sufficiently little.

For perspective, the formulas imply
\[
 \lim_{\alpha\downarrow0}\eta_H(\alpha)=\frac35,
 \qquad
 \lim_{\alpha\downarrow0}\eta_C(\alpha)=\frac23.
\]
At the other extreme, expansion of the integrands near $t=1$ gives
\[
 R_H(\alpha)\sim\frac{6e^\alpha}{\alpha^3},\qquad
 R_C(\alpha)\sim\frac{4e^\alpha}{\alpha^2},\qquad
 W_H(\alpha)\sim W_C(\alpha)\sim\frac{2e^\alpha}{\alpha}.
\]
Thus $\eta_H(\alpha)\sim3/\alpha^2$ and
$\eta_C(\alpha)\sim2/\alpha$. These conclusions concern the matched
two-bidder models defined here; the signs of cross-signal effects alone
do not imply a general revenue or extraction-ratio ranking.

    \section{Conclusion}
    Our paper and \cite{chen2026winnersblisscommonvalueauctions} appear to be the first to study auctions with interdependent values where bidders' information and uncertainty are differentiated horizontally rather than vertically. Purely horizontal differentiation, as in our baseline model, is admittedly rare in real-world auctions. Perhaps the cleanest example is the oil-tract bidding we discussed in the introduction, but many important auction environments combine both horizontal and vertical differentiation among bidders.

The general co-existence of horizontal and vertical differentiation means that no single existing model fully captures reality. The standard interdependent-values framework in the literature, what we have called the winner's curse model,  rules out the possibility that good news for one bidder could simultaneously be bad news for others. Yet many environments clearly contain at least some flavor of winner's bliss, where exactly that kind of asymmetry in information and payoffs arises. For example, whenever bidders are invested in distinct technologies and the object for sale has a yet-unknown better fit with some technologies than with others, we would expect some flavor of horizontal differentiation and winners' bliss.

We hope that our model and paper serve as motivation for a more systematic analysis of hybrid environments: models that blend both winner’s curse and winner’s bliss elements, and thereby come closer to the complexity of real-world auctions.
\clearpage
    \appendix

\section{Proof of \cref{thm:english}}\label{pf:english_thm}

\begin{proposition}\label{prop:eqm}
    Suppose a valuation $\mathbf{v}$ satisfies OED and $ \max_{i \in N} v_i(\mathbf{0})< \min_{i \in N} v_i(\mathbf{1})$. Then for all $A \subseteq N$, for all $\mathbf{x}_{N \setminus A}$, there exists a unique set of continuous and increasing functions $\bm{\sigma}_A = (\sigma_i)_{i\in A}$, where $$\sigma_i:  \left[ \max_{i \in A} v_i(\mathbf{0},\mathbf{x}_{N \setminus A}), \min_{i \in A} v_i(\mathbf{1},\mathbf{x}_{N \setminus A}) \right] \to [0,1],$$ such that $v_i(\bm{\sigma}_A(p),\mathbf{x}_{N \setminus A}) = p$,  for all $p \in  \left[ \max_{i \in A} v_i(\mathbf{0},\mathbf{x}_{N \setminus A}), \min_{i \in A} v_i(\mathbf{1},\mathbf{x}_{N \setminus A}) \right]$.
      \begin{proof}

      We will start by proving the existence of $\bm{\sigma}_A$, then we will prove the monotonicity.

    \begin{lemma}\label{lemma:implicit}
          Suppose a valuation $\mathbf{v}$ satisfies OED and $ \max_{i \in N} v_i(\mathbf{0})< \min_{i \in N} v_i(\mathbf{1})$. Then for all $A \subseteq N$, for all $\mathbf{x}_{N \setminus A}$, there exists a set of continuous functions $\bm{\sigma}_A = (\sigma_i)_{i\in A}$, where $\sigma_i:  \left[ \max_{i \in A} v_i(\mathbf{0},\mathbf{x}_{N \setminus A}), \min_{i \in A} v_i(\mathbf{1},\mathbf{x}_{N \setminus A}) \right] \to [0,1]$, such that $v_i(\bm{\sigma}_A(p),\mathbf{x}_{N \setminus A}) = p$,  for all $p \in  \left[ \max_{i \in A} v_i(\mathbf{0},\mathbf{x}_{N \setminus A}), \min_{i \in A} v_i(\mathbf{1},\mathbf{x}_{N \setminus A}) \right]$.    
         \begin{proof}
   Given that $ \max_{i \in N} v_i(\mathbf{0})< \min_{i \in N} v_i(\mathbf{1})$, we first show that the domain of each $\bm{\sigma}_A$ function exists. For all $A \subseteq N$, and $\mathbf{x}_{N \setminus A}$, we have 
   \[
   \max_{i \in A} v_i(\mathbf{0},\mathbf{x}_{N \setminus A}) \le \max_{i \in N} v_i(\mathbf{0})< \min_{i \in N} v_i(\mathbf{1}) \le  \min_{i \in A} v_i(\mathbf{1},\mathbf{x}_{N \setminus A})
   \]

   The proof is based on the implicit function theorem. Define the auxiliary function $\mathbf{F}: [0, 1]^{|A|} \times \mathbb{R} \to \mathbb{R}^{|A|}$:
$$
\mathbf{F}(\mathbf{x}_A, p) = \mathbf{v}(\mathbf{x}_A,\mathbf{x}_{N \setminus A}) - p \cdot\mathbf{1} 
$$
where $F_i(\mathbf{x}_A, p) = v_i(\mathbf{x}_A,\mathbf{x}_{N\setminus A}) - p$. Note that $\mathbf{x}_{N \setminus A}$ is given. We first prove the existence of $\mathbf{x}_A \in [0,1]^{|A|}$ such that $\mathbf{F}(\mathbf{x}_A,p) = \mathbf{0}$, for all $p \in  \left[ \max_{i \in A} v_i(\mathbf{0},\mathbf{x}_{N \setminus A}), \min_{i \in A} v_i(\mathbf{1},\mathbf{x}_{N \setminus A}) \right]$. $\mathbf{F}(\mathbf{x}_A,p) = \mathbf{0}$ implies that $
v_i(\mathbf{x}_A,\mathbf{x}_{N \setminus A})  = p
$, for all $i \in A$, which is what we aim to prove.

Note that, for all $p \in  \left[ \max_{i \in A} v_i(\mathbf{0},\mathbf{x}_{N \setminus A}), \min_{i \in A} v_i(\mathbf{1},\mathbf{x}_{N \setminus A}) \right]$, we have
\begin{align*}
    v_j(0,\mathbf{x}_{A \setminus \{j\}},\mathbf{x}_{N \setminus A}) \le v_j(\mathbf{0},\mathbf{x}_{N \setminus A}) \le \max_{i \in A} v_i(\mathbf{0},\mathbf{x}_{N \setminus A}) \le p\\
    v_j(1,\mathbf{x}_{A \setminus \{j\}},\mathbf{x}_{N \setminus A}) \ge v_j(\mathbf{1},\mathbf{x}_{N \setminus A}) \ge \min_{i \in A} v_i(\mathbf{1},\mathbf{x}_{N \setminus A}) \ge p,
\end{align*}
for all $j \in A$. We can apply the Poincaré–Miranda Theorem to show that, for all $$p \in  \left[ \max_{i \in A} v_i(\mathbf{0},\mathbf{x}_{N \setminus A}), \min_{i \in A} v_i(\mathbf{1},\mathbf{x}_{N \setminus A}) \right],$$ there exists $\mathbf{x}_A \in [0,1]^{|A|}$ such that
$
v_i(\mathbf{x}_A,\mathbf{x}_{N \setminus A})  = p
$, for all $i \in A$.

The Poincar\'{e}–Miranda Theorem ensures that for every $p$ in the interval, the level set is non-empty. Moreover, the solution to the system is unique. To see that, assume there exist two vectors $\mathbf{x}_A$ and $\mathbf{x}_A'$ such that $
v_i(\mathbf{x}_A,\mathbf{x}_{N \setminus A})  = v_i(\mathbf{x}_A',\mathbf{x}_{N \setminus A}) = p
$, for all $i \in A$. Select the bidder $j \in A$ such that $|x_j-x_j'| \ge |x_i - x_i'|$, for all $i \in A$. Without loss of generality, let $x_j - x_j' = \epsilon > 0$. Let $D = \{i\in A: x_i-x_i' < 0\}$, and we have
\begin{align*}
    & v_j(\mathbf{x}_A,\mathbf{x}_{N \setminus A})  -v_j(\mathbf{x}_A',\mathbf{x}_{N \setminus A})\\
     =&  v_j(x_j'+\epsilon,\mathbf{x}_{D\setminus \{j\}}, \mathbf{x}_{A \setminus D},\mathbf{x}_{N \setminus A})  -v_j(x_j',\mathbf{x}'_{D\setminus \{j\}}, \mathbf{x}'_{A \setminus D},\mathbf{x}_{N \setminus A})\\
    > & v_j(x_j'+\epsilon,\mathbf{x}'_{D\setminus \{j\}}, \mathbf{x}_{A \setminus D},\mathbf{x}_{N \setminus A})  -v_j(x_j',\mathbf{x}'_{D\setminus \{j\}}, \mathbf{x}'_{A \setminus D},\mathbf{x}_{N \setminus A})\\
    \ge & v_j(x_j'+\epsilon,\mathbf{x}'_{D\setminus \{j\}}, \mathbf{x}_{A \setminus D}' + \mathbf{1} \cdot \epsilon,\mathbf{x}_{N \setminus A})  -v_j(x_j',\mathbf{x}'_{D\setminus \{j\}}, \mathbf{x}'_{A \setminus D},\mathbf{x}_{N \setminus A})\\
    > & v_j(x_j'+\epsilon,\mathbf{x}_{A }' + \mathbf{1} \cdot \epsilon,\mathbf{x}_{N \setminus A} + \mathbf{1} \cdot \epsilon)  -v_j(x_j', \mathbf{x}'_{A},\mathbf{x}_{N \setminus A}) > 0
\end{align*}

The first inequality comes from the fact that $\mathbf{x}_{D \setminus \{j\}} < \mathbf{x}_{D \setminus \{j\}}'$, and $v_j$ is decreasing in $x_i$, for all $i \ne j$. The second inequality comes from the fact that $|x_j-x_j'| \ge |x_i - x_i'|$, for all $i \in A$, and hence, $\mathbf{x}_{A \setminus D} < \mathbf{x}_{A \setminus D}' + \mathbf{1} \cdot \epsilon$. Finally, the last inequality comes from the OED assumption. This contradicts the assumption that $
v_i(\mathbf{x}_A,\mathbf{x}_{N \setminus A})  = v_i(\mathbf{x}_A',\mathbf{x}_{N \setminus A}) = p
$, which proves the uniqueness of the solution.

The Implicit Function Theorem then allows us to parameterize this solution set as a smooth curve $\bm{\sigma}_A(p)$. To apply the Implicit Function Theorem, there are two more conditions to check. First, we need to check that $\mathbf{F}$ is $C^1$, which is true since we assume that $v_i$ is $C^2$. Second, we need to check that the relevant Jacobian matrix with respect to $\mathbf{x}_A$, denoted by $J(\mathbf{x}_A)$ is not singular, where $J_{ij}(\mathbf{x}_A) = \frac{\partial v_i}{\partial x_j}(\mathbf{x}_A,\mathbf{x}_{N \setminus A})$.

Since $\frac{\partial v_i}{\partial x_j} < 0$ for all $j \neq i$, we have $\left| \frac{\partial v_i}{\partial x_j} \right| = - \frac{\partial v_i}{\partial x_j}$. Combined with the OED condition, this implies
\begin{align*}
    & \sum_{j\in N} \frac{\partial{v}_i}{\partial x_j}> 0 \\
    \implies & \left| \frac{\partial v_i}{\partial x_i} \right| > \sum_{j \neq i} \left| \frac{\partial v_i}{\partial x_j} \right| \\
    \implies & \left| J_{ii}(\mathbf{x}_A) \right| > \sum_{j \neq i} \left| J_{ij} (\mathbf{x}_A) \right| \ \ \forall \ \mathbf{x}_A \in [0,1]^{|A|}.
\end{align*}

We can apply the Levy-Desplanques Theorem, and claim that $\det(J(\mathbf{x}_A)) \ne 0$, for all solutions $(\mathbf{x}_A,p)$ such that $\mathbf{F}(\mathbf{x}_A,p) = 0$. By the implicit function theorem, we can prove that there exists a set of continuous functions $\bm{\sigma}_A = (\sigma_i)_{i\in A}$, where $\sigma_i:  \left[ \max_{i \in A} v_i(\mathbf{0},\mathbf{x}_{N \setminus A}), \min_{i \in A} v_i(\mathbf{1},\mathbf{x}_{N \setminus A}) \right] \to [0,1]$, such that $v_i(\bm{\sigma}_A(p),\mathbf{x}_{N \setminus A}) = p$,  for all $p \in  \left[ \max_{i \in A} v_i(\mathbf{0},\mathbf{x}_{N \setminus A}), \min_{i \in A} v_i(\mathbf{1},\mathbf{x}_{N \setminus A}) \right]$.
         \end{proof}
         \end{lemma}

Finally, we want to show that $\bm\sigma_A$ is increasing. Given a set of functions $\bm\sigma_A = (\sigma_i)_{i \in A}$ constructed by \cref{lemma:implicit}, assume that there exists a $p, \ p' \in \left[ \max_{i \in A} v_i(\mathbf{0},\mathbf{x}_{N \setminus A}), \min_{i \in A} v_i(\mathbf{1},\mathbf{x}_{N \setminus A}) \right]$ and $p<p'$ such that $\sigma_i(p) \ge \sigma_i(p')$ for some $i \in A$. Without loss of generality, let $D = \{1,\ldots,k\} \subseteq A$ and $\sigma_i(p) \ge \sigma_i(p')$ for all $i \in D$.  
         
          By the previous result, we have $v_i(\sigma_1(p), \sigma_2(p),\ldots, \sigma_n(p)) = p $ and $v_i(\sigma_1(p'), \sigma_2(p'),\ldots, \sigma_n(p')) = p' $ for all $1 \le i \le n$. Hence, we have
         \begin{align*}
             & v_1(\sigma_1(p'), \sigma_2(p'),\ldots, \sigma_n(p')) - v_1(\sigma_1(p), \sigma_2(p),\ldots, \sigma_n(p)) \\
             = & \sum_{i=1}^n v_1(\sigma_1(p'),\dots, \sigma_i(p'),\sigma_{i+1}(p),\ldots, \sigma_n(p)) - v_1(\sigma_1(p'),\dots, \sigma_i(p),\sigma_{i+1}(p),\ldots, \sigma_n(p)) > 0
         \end{align*}

         Now, we first claim that there exists $j \in D$ such that $\sigma_j(p) - \sigma_j(p') >\sigma_1(p) - \sigma_1(p')$ (note that $\sigma_i(p) \ge \sigma_i(p')$ for all $i \in D$). Assume that this claim is false, that is, $\sigma_1(p) - \sigma_1(p') \ge \sigma_j(p) - \sigma_j(p')$, for all $j \in D$. Let $\sigma_1(p) - \sigma_1(p') = \epsilon \ge 0$, we have
        \begin{align*}
             & \sum_{i=1}^n v_1(\sigma_1(p'),\dots, \sigma_i(p'),\sigma_{i+1}(p),\ldots, \sigma_n(p)) - v_1(\sigma_1(p'),\dots, \sigma_i(p),\sigma_{i+1}(p),\ldots, \sigma_n(p)) \\
             < & \sum_{i=1}^k v_1(\sigma_1(p'),\dots, \sigma_i(p'),\sigma_{i+1}(p),\ldots, \sigma_n(p)) - v_1(\sigma_1(p'),\dots, \sigma_i(p),\sigma_{i+1}(p),\ldots, \sigma_n(p))\\
             = &  v_1(\sigma_1(p'),\dots, \sigma_k(p'),\sigma_{i+1}(p),\ldots, \sigma_n(p)) - v_1(\sigma_1(p),\ldots, \sigma_n(p))\\
             \le & v_1(\sigma_1(p)-\epsilon,\dots,\sigma_k(p)-\epsilon,\sigma_{k+1}(p),\dots,\sigma_n(p))- v_1(\sigma_1(p),\dots,\sigma_n(p)) \\
             < & v_1(\sigma_1(p)-\epsilon,\dots,\sigma_n(p)-\epsilon) - v_1(\sigma_1(p),\dots,\sigma_n(p)) < 0 
         \end{align*}

        The first inequality comes from the fact that $\sigma_i(p') > \sigma_i(p)$ and $v_1$ is decreasing in $x_i$ for all $i > k$. Hence, we have
        \begin{align*}
          & v(\sigma_1(p'),\dots,\sigma_i(p'),\sigma_{i+1}(p),\dots,\sigma_n(p)) - v(\sigma_1(p'),\dots,\sigma_i(p),\sigma_{i+1}(p),\dots,\sigma_n(p)) < 0, \ \forall \ i>k \\ 
         \implies & \sum_{i=k+1}^n v_1(\sigma_1(p'),\dots, \sigma_i(p'),\sigma_{i+1}(p),\ldots, \sigma_n(p)) - v_1(\sigma_1(p'),\dots, \sigma_i(p),\sigma_{i+1}(p),\ldots, \sigma_n(p))< 0
        \end{align*}

        And the second inequality comes from the assumption that  $ \epsilon = \sigma_1(p) - \sigma_1(p') \ge \sigma_i(p) - \sigma_i(p')$ and $v_1$ is decreasing in $x_i$ for all $ 1 < i \le k$. Hence, we have $\sigma_i(p') \ge \sigma_i(p) - \epsilon.$ Finally, the last inequality comes from the OED assumption.
         
        This result will give us a contradiction to the assumption that $p' > p$. Hence, there must exist a  $j \in D$ such that $\sigma_j(p) - \sigma_j(p') >\sigma_1(p) - \sigma_1(p')$. If we replace $1$ with any other $i\le k$, this result will also hold. Therefore, we conclude that for all $i \in D$ there exists $j \in D$ such that $\sigma_j(p) - \sigma_j(p') >\sigma_i(p) - \sigma_i(p')$, which contradicts the fact that $D$ is a finite set.
    \end{proof}
\end{proposition}

Now, we construct a bidding strategy $ \beta_i^A(x_i,\mathbf{p}_{N\setminus A})$, which can also be written as $\beta_i^A(x_i,\mathbf{x}_{N\setminus A})$. This is because once a bidder drops out at some price, all the remaining bidders can deduce the signal she received via her bidding strategy. With the result in \cref{lemma:implicit}, we define $\beta_i^A(x_i,\mathbf{x}_{N \setminus A}) = \min \{ p: \sigma_i(p) \ge x_i\}$, where $\bm\sigma_A$ is constructed with respect to $\mathbf{x}_{N \setminus A}$. Let $\bm\beta = ((\beta_i)_{i \in A})_{A \subseteq N}$ denote the bidding strategy profile.

\begin{proposition}
    Suppose a valuation $\mathbf{v}$ satisfies OED and $ \max_{i \in N} v_i(\mathbf{0})< \min_{i \in N} v_i(\mathbf{1})$. Then the bidding strategy profile $\bm\beta = ((\beta_i)_{i \in A})_{A \subseteq N}$ constitutes an efficient equilibrium under the English auction.
    \begin{proof}
        We first invoke the result from \textcite{krishna2003asymmetric}, which states that \cref{lemma:implicit} implies that $\bm\beta$ sustains an ex post equilibrium of the English auction. We then prove the efficiency of this equilibrium.

        Assume that the realized signal profile is $\mathbf{x} = (x_1, \dots, x_n)$. Let $p_k$ denote the dropout price of bidder $k$, given that they follow the strategy profile $\bm\beta$ (where for the winner of the auction, this is the price at which they would have dropped out if other bidder would have stayed in). Without loss of generality, assume that the dropout prices are ordered such that $p_1 > p_2 > \dots > p_n$, implying that Bidder 1 is the winner. The goal is to show that the winner (Bidder 1) has the highest valuation at the true signal profile $\mathbf{x}$, i.e., $v_1(\mathbf{x}) \ge v_j(\mathbf{x})$ for all $j \neq 1$.

        Note that, once a bidder drops out, the signal she received can be deduced due to the structure of the equilibrium, and all the remaining bidders will apply a new $\bm\sigma_A$ function as their bidding strategies going forward. Since we assume that people drop out in descending order, we write $\bm\sigma^{j} = (\sigma_i^j)_{i \le j}$ to denote the function that bidders apply in the time interval where bidder $j+1$ has dropped out but $j$ has not. For any $j > 1$, at the price she plans to drop out, we have   
        \[
        v_1(\sigma_1^{j}(p_j), \ldots, \sigma^{j}_j(p_j), x_{j+1}, \ldots, x_n) = v_j(\sigma_1^j(p_j), \ldots, \sigma_j^j(p_j), x_{j+1}, \ldots, x_n),\] which is simply a consequence of the definition of $\bm\sigma^j$.


        We first show that $v_1(\mathbf{x})>v_2(\mathbf{x})$. At the price $p_2$, we know that 
        \[
        v_1(\sigma_1^2(p_2), \sigma_2^2(p_2), x_3, \ldots, x_n) = v_2(\sigma_1^2(p_2), \sigma_2^2(p_2), x_3, \ldots, x_n)\]
        We also know that
        \begin{align*}
            v_1(\sigma_1^2(p_1), \sigma_2^2(p_2), x_3, \ldots, x_n) & > v_1(\sigma_1^2(p_2), \sigma_2^2(p_2), x_3, \ldots, x_n) \\
            v_2(\sigma_1^2(p_1), \sigma_2^2(p_2), x_3, \ldots, x_n) & < v_2(\sigma_1^2(p_2), \sigma_2^2(p_2), x_3, \ldots, x_n)
        \end{align*}
        since $\sigma_i(p)$ is increasing. Since $x_1 = \sigma_1^2(p_1)$ and  $x_2 = \sigma_2^2(p_2)$, the above implies $v_1(\mathbf{x})>v_2(\mathbf{x})$.

    Now, let's prove the following claim: if $v_1(\mathbf{x})>v_{i}(\mathbf{x})$ for all $i \in \{2, \dots, j-1\}$, then $v_1(\mathbf{x}) > v_{j}(\mathbf{x})$. Assume that the claim is false and there exists $j>1$ such that $v_1(\mathbf{x})>v_{i}(\mathbf{x})$ for all $i \in \{2, \dots, j-1\}$, but $v_1(\mathbf{x}) \le v_{j}(\mathbf{x})$. We define $\alpha^{p_j}_i = v_i(\mathbf{x}) - p_j$. First note that,
    \begin{align*}
        \alpha^{p_j}_j & = v_j(\mathbf{x}) - p_j\\
        & = v_j(x_1,x_2,\dots,x_j,x_{j+1},\dots,x_n) - v_j(\sigma_1^j(p_j),\sigma_2^j(p_j),\dots,\sigma_j^j(p_j),x_{j+1},\dots,x_n) <0
    \end{align*}

    To see that the above is true, first note that $x_j = \sigma_j^j(p_j)$. As for $x_i$ such that $i < j$, we have $x_i > \sigma_i^j(p_j)$, if this is not true, we will have that $i$ drops out at the price lower than $p_j$, which contradicts our assumption. Given that $x_i > \sigma_i^j(p_j)$, for all $i \in \{2, \dots, j-1\}$, and $v_j$ is decreasing in $x_i$ for all $i \ne j$, we conclude that $\alpha_j^{p_j}$ is negative. Combined with the assumption that $v_i(\mathbf{x}) < v_1(\mathbf{x}) \le v_j(\mathbf{x})$, for all $i \in \{2, \dots, j-1\}$, we have $\alpha^{p_j}_i < \alpha^{p_j}_1 \le \alpha^{p_j}_j < 0$, for all $i \in \{2, \dots, j-1\}$.
    
    Let $i \in \{1, \dots, j-1\}$ be the index that maximizes the difference $x_k - \sigma_k^j(p_j)$ among all $k < j$. Let $\epsilon = x_i - \sigma_i^j(p_j)$. Now,
    \begin{align*}
          \alpha_i^{p_j} &=   v_i(x_1,\dots,x_i,\dots,x_j,x_{j+1},\dots,x_n) -  v_i(\sigma_1^j(p_j),\dots,\sigma_i^j(p_j),\dots,\sigma_j^j(p_j),x_{j+1},\dots,x_n) \\ 
           &> v_i(\sigma_1^j(p_j)+\epsilon,\dots,\sigma_i^j(p_j)+\epsilon,\dots,\sigma_j^j(p_j)+\epsilon,x_{j+1},\dots,x_n)\\
           & - v_i(\sigma_1^j(p_j),\dots,\sigma_i^j(p_j),\dots,\sigma_j^j(p_j),x_{j+1},\dots,x_n)\\
            & > v_i(\sigma_1^j(p_j)+\epsilon,\dots,\sigma_j^j(p_j)+\epsilon,x_{j+1}+\epsilon,\dots,x_n+\epsilon)\\
           & -v_i(\sigma^j_1(p_j),\dots,\sigma^j_j(p_j),x_{j+1},\dots,x_n) > 0.
     \end{align*} The first two inequalities come from the assumption that $v_i$ is decreasing in $x_k$, for all $k \ne i$. And the last inequality comes from the OED assumption.

    Now, our earlier derivation established that $\alpha_i^{p_j} < 0$ for all $i < j$. This is a contradiction. Therefore, the assumption that 
    $v_1(\mathbf{x}) \le v_j(\mathbf{x})$ must be false.
    
    We conclude that $v_1(\mathbf{x}) > v_j(\mathbf{x})$. Combining this with the base case 
    $v_1(\mathbf{x}) > v_2(\mathbf{x})$, the result $v_1(\mathbf{x}) > v_k(\mathbf{x})$ 
    holds for all $k > 1$ by induction. Thus, the equilibrium is efficient.
    \end{proof}
\end{proposition}

\printbibliography

@online{chen2026winnersblisscommonvalueauctions,
      title={The Winner's Bliss in Common-Value Auctions under Horizontal Differentiation}, 
      author={Jiawei Chen and Anh Nguyen and Matthew Shum},
      year={2026},
      eprint={2606.08419},
      archivePrefix={arXiv},
      primaryClass={econ.GN},
      url={https://arxiv.org/abs/2606.08419}, 
}

@article{HAILE2010390,
title = {Recent U.S.\ offshore oil and gas lease bidding: A progress report},
journal = {International Journal of Industrial Organization},
volume = {28},
number = {4},
pages = {390-396},
year = {2010},
note = {Selected Papers, European Association for Research in Industrial Economics 36th Annual Conference, Ljubljana, Slovenia, September 3-5, 2009},
issn = {0167-7187},
doi = {https://doi.org/10.1016/j.ijindorg.2010.02.010},
url = {https://www.sciencedirect.com/science/article/pii/S0167718710000287},
author = {Philip Haile and Kenneth Hendricks and Robert Porter}
}

@incollection{maskin92,
    author = {Maskin, Eric S},
    booktitle = {Privatization},
editor = {H. Siebert},
    publisher = {J.C.B. Mohr Publisher ; },
    year = {1992},
    title = {Auctions and privatization}
}

@book{krishna2009auction,
  title={Auction theory},
  author={Krishna, Vijay},
  year={2009},
  publisher={Academic press}
}

@article{dubra2009english,
  title={English auctions and the Stolper--Samuelson theorem},
  author={Dubra, Juan and Echenique, Federico and Manelli, Alejandro M},
  journal={Journal of Economic Theory},
  volume={144},
  number={2},
  pages={825--849},
  year={2009},
  publisher={Elsevier}
}

@article{krishna2003asymmetric,
  title={Asymmetric {E}nglish auctions},
  author={Krishna, Vijay},
  journal={Journal of Economic Theory},
  volume={112},
  number={2},
  pages={261--288},
  year={2003},
  publisher={Elsevier}
}

@article{krishna1997analysis,
  title={An analysis of the war of attrition and the all-pay auction},
  author={Krishna, Vijay and Morgan, John},
  journal={Journal of Economic Theory},
  volume={72},
  number={2},
  pages={343--362},
  year={1997},
  publisher={Elsevier}
}

\end{document}